\documentclass[useAMS,usenatbib]{mn2e}
\usepackage{graphicx}
\usepackage{graphics}

\usepackage{booktabs}
\newcommand{\ra}[1]{\renewcommand{\arraystretch}{#1}}

\title[SFDM mass model and evolution of rotation curves for Lsb galaxies]{Scalar Field Dark Matter mass model and evolution of rotation curves for Lsb galaxies}
\author[L. A. Martinez-Medina and T. Matos]{L. A.~Martinez-Medina,\thanks{E-mail: lmedina@fis.cinvestav.mx}\thanks{Part of the Instituto Avanzado de Cosmolog\'ia (IAC) collaboration http://www.iac.edu.mx/} and T.~Matos,\thanks{E-mail: tmatos@fis.cinvestav.mx}\\
  Departamento de F\'isica, Centro de Investigaci\'on y de Estudios Avanzados del IPN, A.P. 14-740, 07000 M\'exico D.F., M\'exico.}

\date{}

\begin{document}

\label{firstpage}

\maketitle

\begin{abstract} 

We study the evolution of gas rotation curves within the scalar field dark matter (SFDM) model. In this model the galactic haloes are astronomical Bose-Einstein Condensate drops of scalar field. These haloes are characterized by a constant-density core and are consistent with observed rotation curves of dark matter dominated galaxies, a missing feature in CDM haloes resulting from DM-only simulations.
We add the baryonic component to the SFDM haloes and simulate the evolution of the dark matter tracer in a set of grid-based hydrodynamic simulations aimed to analyse the evolution of the rotation curves and the gas density distribution in the case of dark matter dominated galaxies. Previous works had found that when considering an exact analytic solution for a static SF configuration, the free parameters of the model allows for good fits to the rotation curves, we confirm that in our simulations but now taking into account the evolution of the baryonic component in a static dark matter and stellar disk potential. Including live gas is a step forward from the previous work using SFDM, as for example, the rotation velocity of the gas is not always exactly equal to the circular velocity of a test particle on a circular orbit. Contrasting with the data the cored mass model presented here is preferred instead of a cuspy one.
\end{abstract}


\section{Introduction}
\label{sec:Intro}

Observed rotation curves along with a mass model are one of the principal tools to derive the gravitational potential and distribution of mass in galaxies. The Inferred dynamical mass and luminosity profile suggest the presence of a dominant dark matter component in galaxies, but a study of rotation curves can give us a more detailed picture of this dark matter component. 
For long time ago there has been tension between predicted and observed distributions of dark matter \citep{F1994,M1994} and an increasing number of recent high quality observations suggest that, for the dark matter distribution, a core-like behaviour is preferred in the central regions of dwarf Spheroidals (dSphs) \citep{G2006,Battaglia2008,Walter2011} and Low surface brightness (LSB) galaxies \citep{O11,R13,dB01,K03,A12}.

A model that tries to explain the nature of dark matter should be able to reproduce this core-like behaviour in the central region of the dark matter haloes. Cored density profiles predicted by the model can be tested against rotation curves. LSB galaxies and the dSphs of the Milky Way are overwhelmingly dark matter dominated, and so provide natural testing grounds for assessing these predictions.

On the other hand, it is a unique prediction of the Cold Dark Matter (CDM) model that the centers of dark haloes are cusped. 
The finding that the haloes of at least some of the dSphs are cored may be interpreted in one
of two ways. Either it provides evidence that baryonic processes such as star formation and supernovae feedback have modified the pristine dark matter cusp \citep{Read2005,Mashchenko2008,C11,Po12}. Or it provides evidence of a new kind of DM, different from the one proposed by the CDM model that holds out the possibility of resolving some of these issues.
About the first way, is still not conclusive that baryonic processes such as star formation and supernova feedback will be able to erase the cusp predicted by CDM-only simulations \citep{Pe12,K11b}, specially in LSB and dSphs galaxies. For this reason is still open the possibility of an alternative explanation for the DM nature that produces cores naturally.

A dark matter scenario that has received much attention is the scalar field dark matter (SFDM) model. The main idea is that the nature of the DM is completely determined by a fundamental scalar field  \citep{Bohmer2007,G00,M01,M12}. In this dark matter and galaxy formation scenario the DM haloes are naturally cored.
\citet{R13} found good agreement with data of rotation curves for LSB galaxies using the minimum disk hypothesis (neglecting the baryonic component).

Because baryons trace the DM potential, the aim of this work is to go forward within the SFDM model by building a galaxy mass model for dark matter dominated galaxies, including a baryonic disk. And because the rotation velocity of the tracer (gas, stars) is not always exactly equal to the circular velocity of a test particle on a circular orbit we simulate the HI tracer evolving the hydrodynamic equations for a gas distribution embedded in our modeled potential in order to measure the rotation curve directly on the gas and contrast with data.

The paper is organized as follow. In section \ref{sec:LSB} we describe briefly the importance of LSB galaxies and HI as a tracer. In section \ref{sec:massmodel} we present the mass model. Section \ref{sec:Initialconditions} contains the techniques an initial conditions for the simulations. Finally we present the results of our numerical simulations and a discussion.

\section{LSB Galaxies and HI as a dark matter tracer}
\label{sec:LSB}

LSB galaxies share common characteristics: late-type, gas-rich, poor star formation rates (SFR), low metallicities, diffuse stellar disks and extended HI gas disks \citep{Bothun97,Bell2000}. 

Another important feature of LSB galaxies is that the mass-to-luminosity ratio is usually higher than that of a normal spiral galaxy due to the low luminosity, and the dark matter fraction is much higher.

The extended HI gas disk in LSB galaxies, although optically difficult to detect, are easily detected in HI surveys and the HI rotation curves of LSB galaxies have received great attention, with shapes different from those of high surface brightness (HSB) galaxies. The rotation curves of HSB spiral galaxies rise fairly steeply to reach an approximately flat part, whereas for LSB galaxies the rotation curves rise more slowly and often still rising at the outermost measured point. In order to describe the data well this galaxies need a massive dark matter halo with a nearly constant density core \citep{K11a}.

With a dark matter component dominating the total mass and HI gas disks far more extended that the stellar disks, LSB galaxies provide a suitable scenario to test the prediction of our mass model trought the study of the rotation curves of a tracer.

\section{Mass model}
\label{sec:massmodel}

For modeling a LSB galaxy we use an analytic two-component galaxy model which is composed of an axisymmetric disk and a spherically symmetric dark matter halo.

The gas and stellar components are represented by a three dimensional Miyamoto-Nagai disk profile \citep{MN75} that is mathematically simple, with a closed expression for the potential

\begin{equation}
\phi_d(R,z) = \frac{-GM_d}{\sqrt{R^2+(a+\sqrt{z^2+b^2})^2}},
\end{equation}

where $M_d$ is the total mass of the component, and $a$ and $b$ are the radial and vertical scale-length, respectively. This potential has continuous derivatives that make it particularly suitable for numerical work.

For the dark matter halo we made our choice based in numerous studies of rotation curves of LSB galaxies that have found the data to be consistent with a halo having a nearly constant density core \citep{dB01,K08,dB10,K11a}, and in the scalar field dark matter model (SFDM) the halos of galaxies featured a constant density core that comes out naturally in the model instead of just been assumed to fit the data.

Aimed to resolved some discrepancies when just taking in to account the condensed system at temperature $T=0$, \citet{R13} consider an scenario in which the dark matter, an auto-interacting real scalar field, is embedded in a thermal bath at temperature $T$. At high temperatures in the early universe, the system interacts with the rest of the matter. Its temperature will decrease  due to the expansion of the universe and eventually, when the temperature is sufficiently small, the scalar field decouples from the interaction with the rest of the matter and follows its own thermodynamic history.

The Newtonian limit provides a good description at galactic scales so there is an exact analytic solution in the Newtonian approximation of this system \citep{R13}, a density profile that accounts for SFDM halos in condensed state or halos in a combination of exited states, $\rho = \rho_0\sin^2(r/a_h)/(r/a_h)^2$, where $\rho_0$ and $a_h$ are fitting parameters.

So the potential of the DM distribution can be written as 

\begin{equation}
\label{eq:DMpotential}
\phi_h(r)=\frac{GM_0}{f(r_0)}\left( \ln(r)+\frac{\sin(2r/a_h)}{2r/a_h}-Ci(2r/a_h)\right)
\end{equation}

where $M_0$ is the mass enclosed within a radius $r_0$, $f(r_0)$ is a constant that depends on $r_0$ and $Ci$ is the cosine integral function. 

For comparison we also use a different dark matter gravitational potential, a NFW halo that results from CDM simulations and characterized by a density profile that rise steeply towards the center \citep{N10}

\begin{equation}
\label{eq:NFW}
\rho_{NFW}(r) = \frac{\rho_s}{(r/r_s)(1+r/rs)^2}
\end{equation}

with $\rho_s$ and $r_s$ as fitting parameters. The gravitational potential from this cuspy halo can be written

\begin{equation}
\label{eq:NFW}
\phi_h(r) = \frac{-GM_0}{g(r_0)r}ln(1+r/r_s)
\end{equation}

with $M_0$ the mass enclosed within a radius $r_0$ and $g(r_0)$ a constant that depends on $r_0$.

The total gravitational potential for our galaxy model is then

\begin{equation}
\label{eq:potential}
\phi(R,z) = \phi_d(R,z) + \phi_h(R,z) ,
\end{equation}

from which we can compute, as usual, the circular velocity of a test particle.

But because the rotation velocity of the gas, that acts as a potential tracer, is not always exactly equal to the circular velocity of a test particle, we ran some simulations evolving the hydrodynamic equations (we do not model star formation or feedback) for a gas distribution embedded in our mass model potential. With this we are able to measure the rotation curves directly on the gas an compare with the data. 
For the data we chose a set of four bulgeless LSB galaxies \citep{McGaugh01,K11b,K08}, because not including a bulge in our mass model reduce the number of adjustable parameters and because for this galaxies the total mass, the stellar mass and the HI mass are known.

For this data catalogue spectra was obtained for a few galaxies in order to check for the presence of noncircular motions, which apparently are minimum in LSB galaxies, as suggested by the authors of the study. Inclinations are also taken into account and errors regarding this only affect the absolute scale of rotation velocities, and not the shape of the rotation curve.

\section{Initial conditions}
\label{sec:Initialconditions}

To measure and study the temporal evolution of the baryonic component under the gravitational potential of our mass model we set up the initial condition of each LSB galaxy as an isolated Miyamoto-Nagai gas disk with a density profile

\begin{equation}
\label{eq:densityprofile}
\rho_d = \frac{b^2M_{HI}}{4\pi}\frac{aR^2+(a+3\sqrt{z^2+b^2})(a+\sqrt{z^2+b^2})^2}{(R^2+(a+\sqrt{z^2+b^2})^2)^{5/2}(z^2+b^2)^{3/2}}.
\end{equation}

where $M_{HI}$ is the HI mass of the galaxy disk, $a$ and $b$ are the radial and vertical scale-length, respectively.

\subsection{Dynamical stability of a gas disk}

Setting up a full three-dimensional dynamically stable astrophysical disk configuration is necessary in order to ensure that the evolution seen in the simulation is not because of some relaxation or evolution towards a new equilibrium configuration driven by non-stable initial conditions. Rotational forces, gravity and velocity dispersions must be balanced in order for the disk does not immediately disperse, collapse or deviate away from the desired density profile. We attempt to construct such a system following the epicyclic approximation as follow.

First we set up the density profile (eq \ref{eq:densityprofile}) and compute the principal circular velocity. Then according to the local stability criterion \citep{Toomre64} we obtain the velocity dispersion in $R$ and $z$ necessary for the disk to be stable against axisymmetric perturbations and be supported by random motions and rotation.

The radial and azimuthal dispersions are given by

\begin{equation}
\label{eq:radialdispersion}
\sigma_R = \frac{3.358\Sigma Q}{K},
\end{equation}

\begin{equation}
\label{eq:radialdispersion}
\sigma_{\phi} = \frac{\sigma_R K}{2\Omega},
\end{equation}

with $\Sigma$ the mass surface density, $Q$ the stability parameter whose initial value is $Q=1.2$ over the entire disk, $\Omega = \sqrt{(\partial_R \phi)/R}$ the angular frequency, and $K$ the epicyclic frequency given by

\begin{equation}
\label{eq:epicyclicfrecquency}
K = \left( 4\Omega^2 + R \frac{d\Omega^2}{dR} \right)^{1/2}.
\end{equation}

And for the vertical velocity dispersion

\begin{equation}
\label{eq:verticaldispersion}
\sigma_z = \sqrt{c\pi G\Sigma b},
\end{equation}

with $c$ a constant and $b$ the vertical scale-length. Finally the mean azimuthal velocity , corrected by non-circular motions (asymmetric drift) is given by \citep{Binney87}

\begin{equation}
\label{eq:asymmetricdrift}
\langle v_{\phi} \rangle^2 = v_c^2 - \sigma_{\phi}^2 -\sigma_R^2( -1 - \frac{R}{\Sigma}\partial_R(\Sigma\sigma_R^2) ).
\end{equation}

The velocity dispersions and the mean azimuthal velocity are then used to distribute the gas component in the velocity space.

\subsection{The code}

The simulations were performed using the latest version of the ZEUS-MP code. It is a fixed-grid, time-explicit Eulerian code that uses an artificial viscosity to handle shocks.

The physics suite in this code includes gas hydrodynamics, ideal MHD, flux-limited radiation diffusion, self gravity, and multispecies advection \citep{H06}.
In this work we focus in solving only the standard hydrodynamics equations for the baryonic component, where the description of the physical state of a fluid element is specified by the following set of fluid equations relating the mass density ($\rho$), velocity ($\textbf{v}$) and gas internal energy density ($e$).

\begin{eqnarray}
\label{eq:2e}
\frac{D\rho}{Dt} + \rho\nabla\cdot\textbf{v} &=& 0,\\
\label{eq:2i}
\rho\frac{D\textbf{v}}{Dt} &=& -\nabla P - \rho\nabla\phi,\\
\label{eq:energia4}
\rho\frac{D}{Dt}\left(\frac{e}{\rho}\right) &=& -P\nabla\cdot\textbf{v},
\end{eqnarray}

where the Lagrangean (or comoving) derivative is given by the usual definition:

\begin{equation}
\label{eq:2l}
\frac{D}{Dt} \equiv \frac{\partial}{\partial t} + \textbf{v}\cdot\nabla.
\end{equation}

The two terms on the right-hand side of the gas momentum equation (\ref{eq:2i}) denote forces due to thermal pressure gradients, and the external gravitational potential, respectively. The self gravity of the gas is neglected in our simulations.
Because most of the material in an astrophysical system is in a highly compressible gaseous phase at very low densities, a perfect adiabatic gas is a good approximation for this system. Here we assume an adiabatic index of $\gamma = 5/3$.

We set the gravitational constant $G$ to unity and the units of mass and length to $M_u = 2.32$ $\times10^7 \rm{M_\odot}$ (galactic mass units) and $R_u = 1 \rm{kpc}$, respectively. With this choice, the resulting units of time and velocity become $t_u = 1 \times 10^8 \rm{yr}$ and $v_u = 10 \rm{kms^{-1}}$.
Cylindrical coordinates $(R; \phi; z)$ are used.

\section{Results}
\label{sec:results}

In this section we describe the temporal evolution of the rotation curves of the gas distribution embedded in the mass models showed above, formed by static dark matter and stellar disk potentials. Then to compare with observations we show the rotation curve measured from the simulation of the isolated gas distribution within the mass model that better fits the data for each one of the LSB galaxies. 
Our SFDM + disk mass model is also compared with the NFW + disk mass model.
The velocity profiles showed next are identical to lower resolution studies for the same parameters, indicating convergence in the results.

\begin{figure}
\begin{center}
\includegraphics[width=8.7cm]{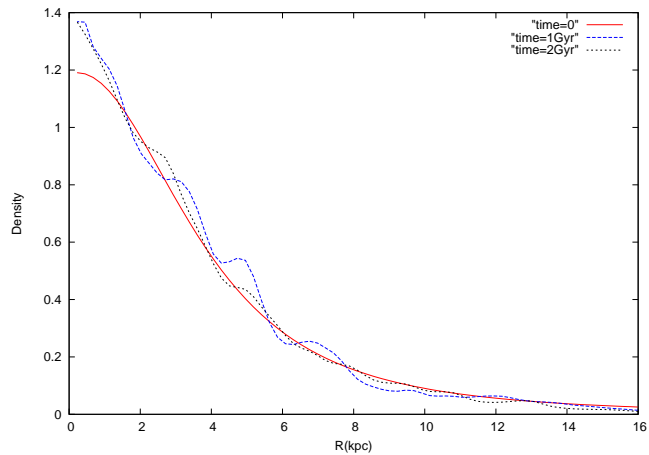}
\end{center}
\caption{Radial density profile for the gas distribution embedded in our mass model, the profiles taken at times $0$, $1Gyr$ and $2Gyr$ }
\label{fig:density}
\end{figure}

But first we tests our initial conditions against dynamical stability, Figure \ref{fig:density} shows the radial density profile for the simulated gas distribution  at different times in the simulation, the gas density profile is evolving but basically oscillating around the Miyamoto-Nagai profile, as desired.

\begin{figure*}
\begin{center}$
\begin{array}{ccc}
\vspace{-0.3cm} 
\includegraphics[width=6cm]{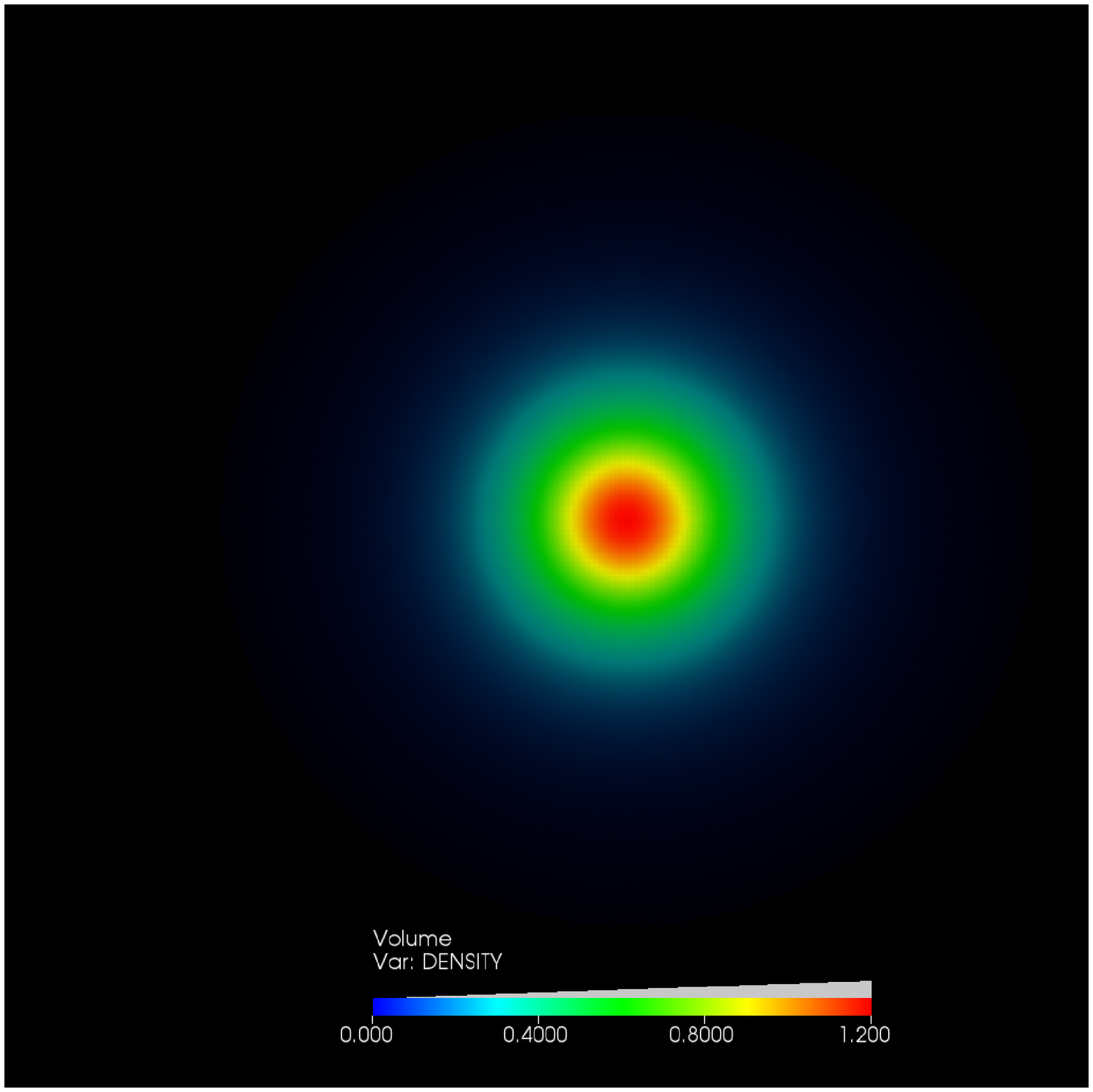} & \hspace{-1cm} \includegraphics[width=6cm]{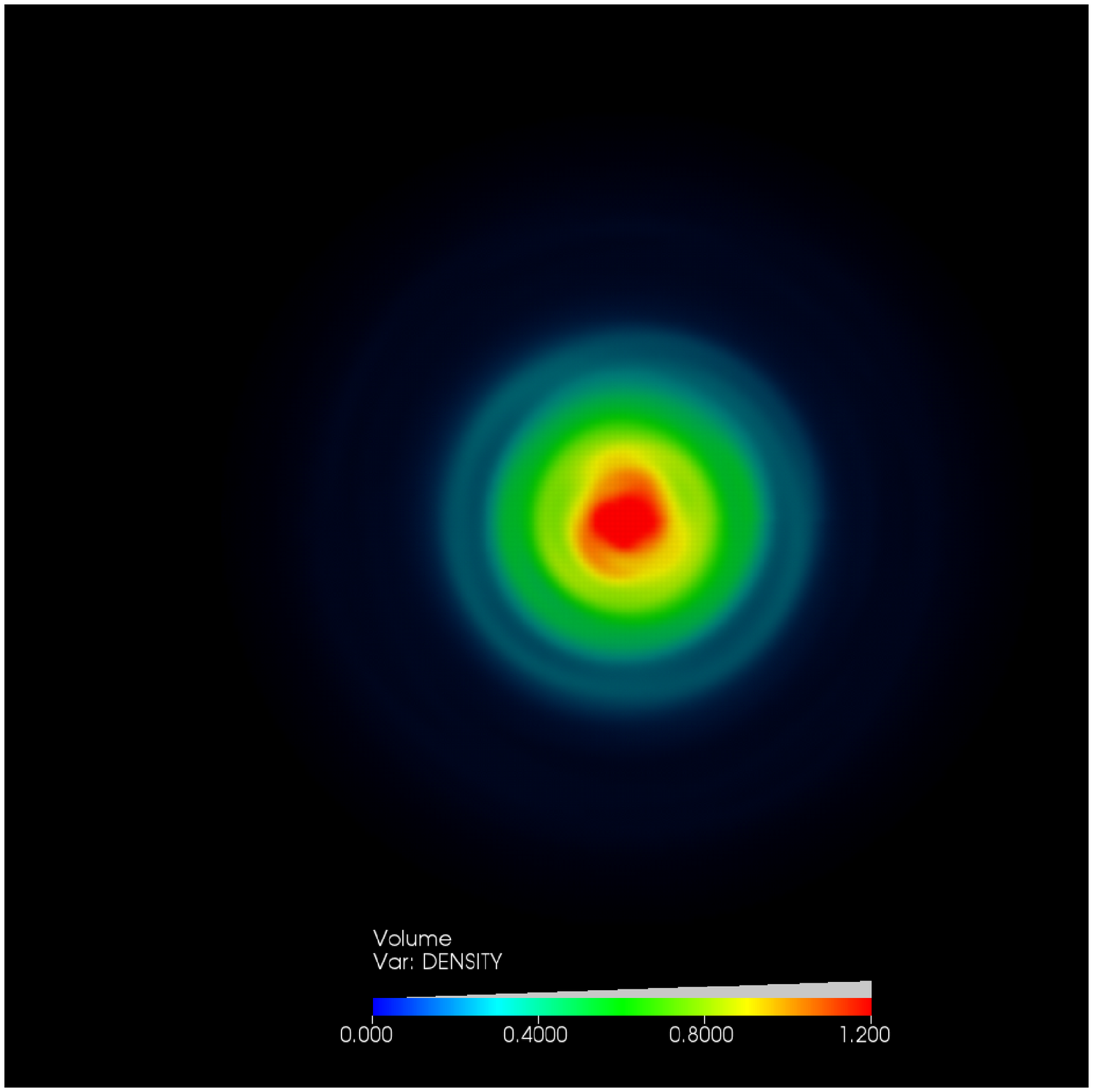}& \hspace{-1cm} \includegraphics[width=6cm]{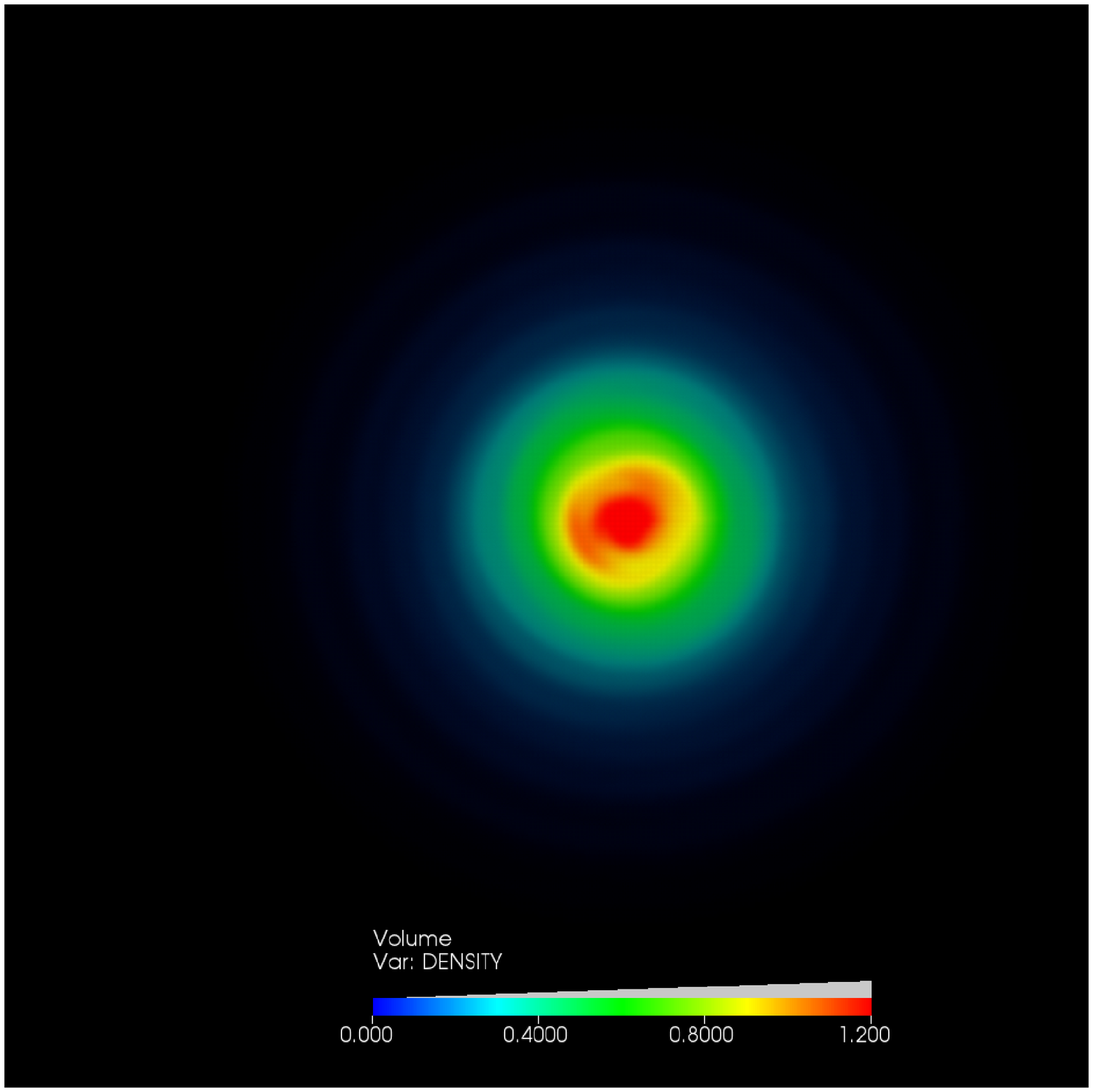}
\end{array}$
\end{center}
\caption{Temporal evolution of the gas density, sliced in the midplane. From left to right, the snapshots show the simulation at: t = 0, 1 Gyr and 2 Gyr.
}
\label{fig:density2}
\end{figure*}

Figure \ref{fig:density2} shows the gas density sliced in the midplane. As seen in this color map, the density do not increase or diminish after a $2$ $\rm{Gyr}$ evolution, indicating stability in the initial conditions.

\begin{figure}
\begin{center}$
\begin{array}{c}
\vspace{-3.cm}
\includegraphics[width=6cm]{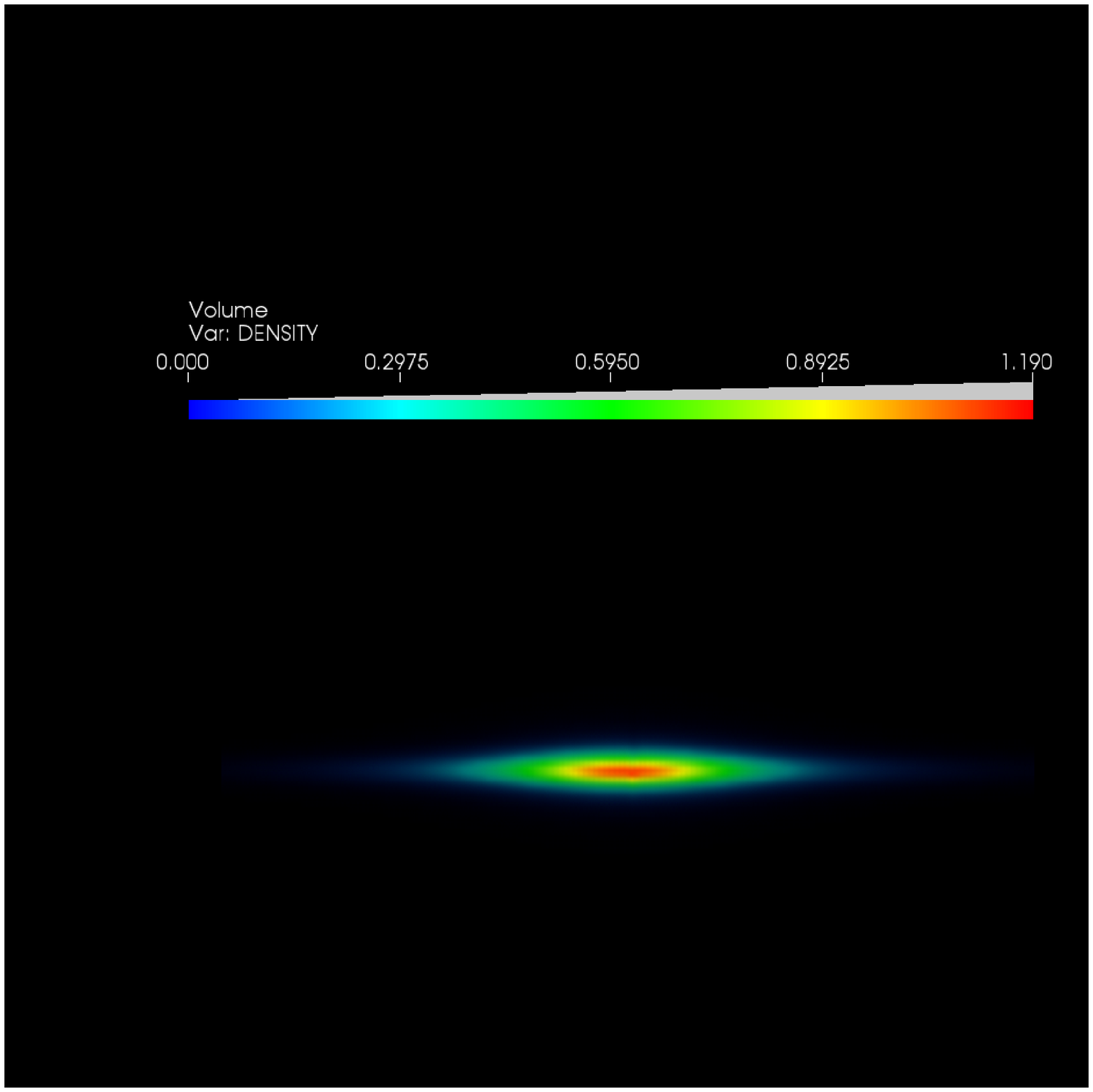}\\
\vspace{-3.cm}
\includegraphics[width=6cm]{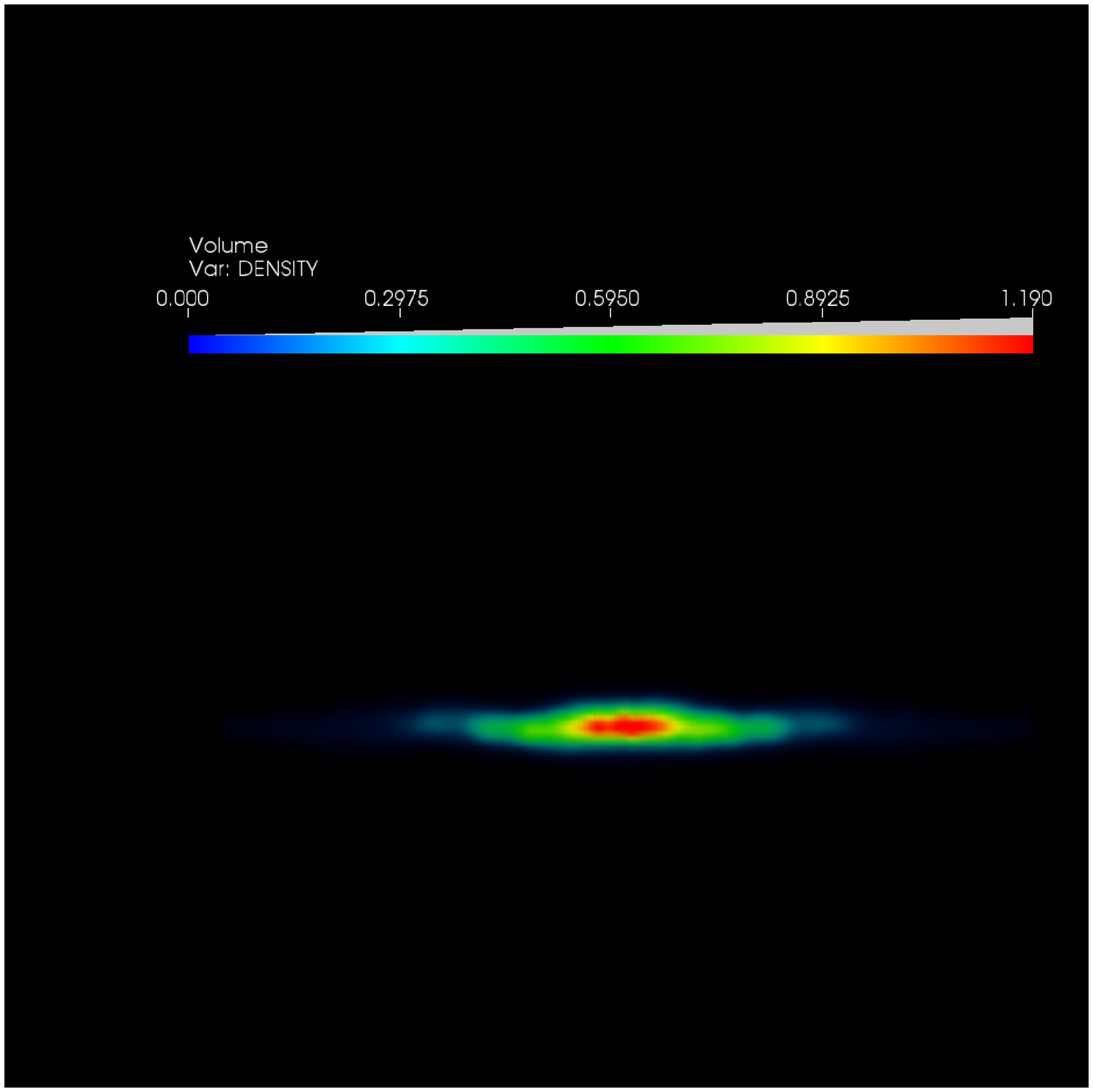}\\
\includegraphics[width=6cm]{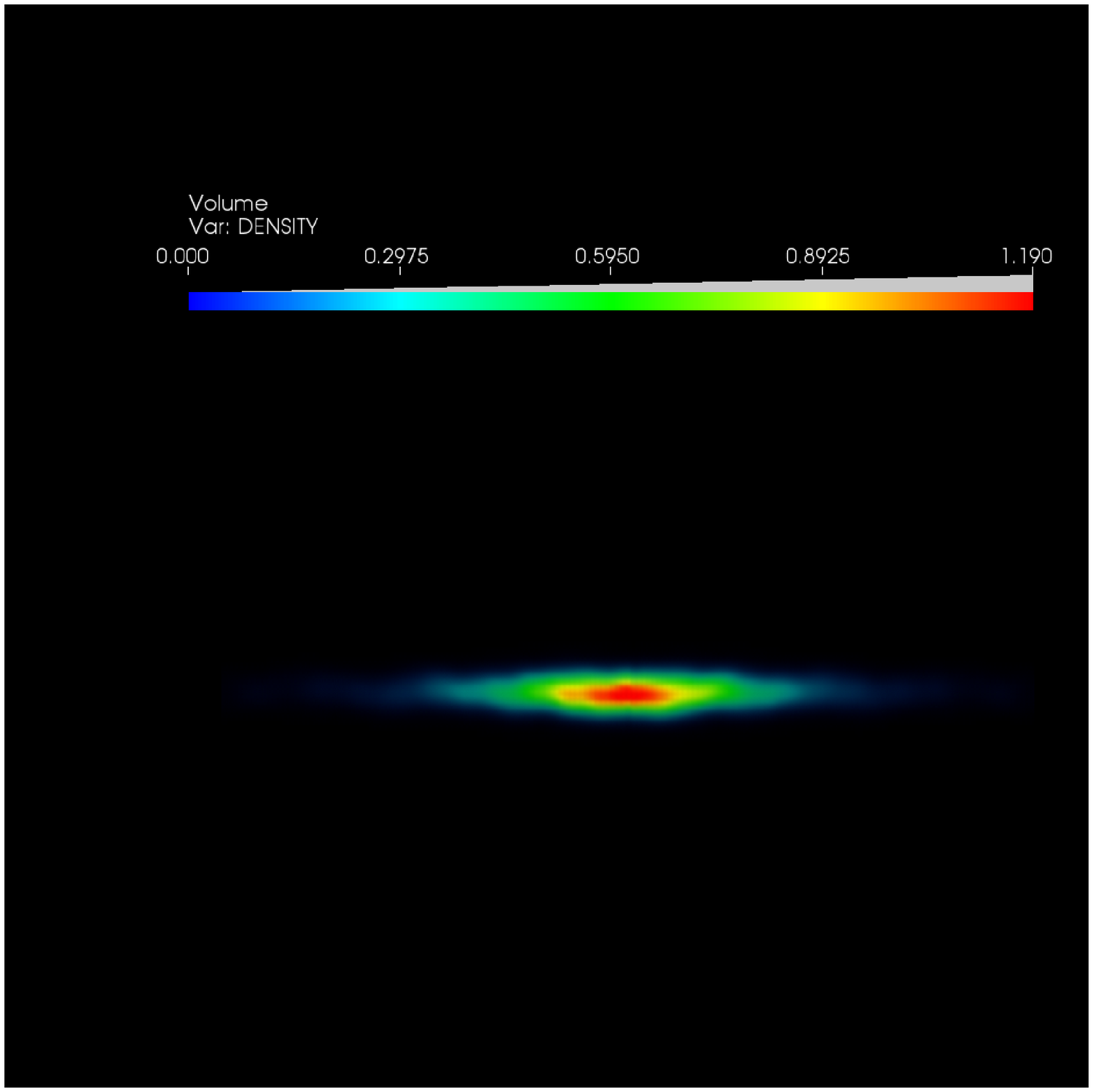}
\end{array}$
\end{center}
\caption{Temporal evolution of the gas density, sliced edge-on From top to bottom the snapshots show the simulation at: t = 0, 1 Gyr and 2 Gyr.
}
\label{fig:density3}
\end{figure}

Meanwhile Figure \ref{fig:density3} plots the edge-on projection of the gas distribution and shows roughly that the density profile do not deviate from the Miyamoto-Nagai during a $2$ \rm{$Gyr$} evolution, there is not spreading or collapsing, just oscillations around the initial configuration.

With this test we are sure that the gas distribution is not contracting, spreading or evolving towards a different profile and that the evolution seen in rotation curves obtained from simulations is not because of a drastic relaxation, as often occurs for artificial initial conditions.

\subsubsection{Time evolution of the rotation curve}

\begin{figure}
\begin{center}
\includegraphics[width=8.7cm]{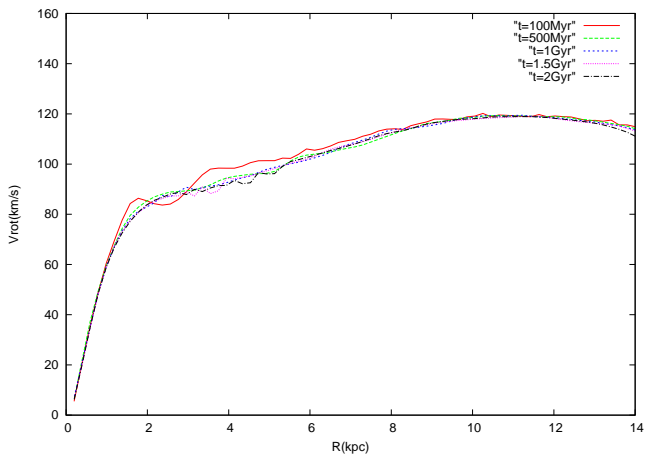}
\end{center}
\caption{Temporal evolution of the rotation curve of the gas distribution embedded in one of our mass models.}
\label{fig:timeevolution}
\end{figure}

The are many physical processes that affect a gas distribution in the galaxy. With evolving just the standard hydrodynamic equations we account for processes that may induce non-circular motions such as velocity gradients, pressure gradients and even a flared or warped gas disk, often seen in HI disks. Figure \ref{fig:timeevolution} shows the temporal evolution of the rotation curve for one of our modeled galaxies evolved for some $2$ \rm{$Gyr$ }.

Processes related to the physics of the tracer, like those mentioned above, could account for wiggles or oscillations seen in the rotation curves. And because they evolve with time is important to consider when trying to confront a mass model with rotation curves data.
From figure \ref{fig:timeevolution} we can see that the rotation curve measured directly on the gas develops wiggles but only at early times in the simulation, this feature smooths quickly and at the end do not deviate that much from the analytical circular velocity.

\subsection{LSB galaxies data}

By performing a manual grid search we locate the parameters for our mass model that represent best four specific, bulgeless, gas-rich, lsb galaxies \citep{McGaugh01,K11b} and contrast the rotation curves data with that obtained from our simulations.

\begin{figure}
\begin{center}
\includegraphics[width=8.7cm]{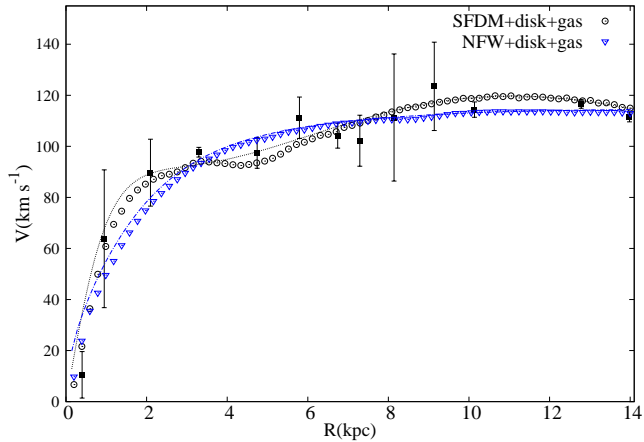}
\end{center}
\caption{F579-V1: Compared Rotation curve from data (square dots), simulation with the SFDM halo + disk (open circles), simulation with the NFW halo + disk (triangles), and the initial rotation curve for both models (dashed lines).}
\label{fig:F579V1}
\end{figure}

F579V1: The SFDM + disk mass model for this Lsb galaxy fits better the data with parameters $a=1.3$ \rm{$kpc$}, $b=0.2$ \rm{$kpc$} for the disk and $a_h=5.2$ \rm{$kpc$} for the dark matter halo. Figure \ref{fig:F579V1} shows the rotation curve resulting from the temporal evolution contrasted with the data. By adjusting only this three parameters we get a relatively good fit, $\chi_r^2=1.3$. 

In order to do a full comparison we perform the same hydrodynamics simulations evolving the gas within the NFW + disk mass model that better fits the data. The constrictions to this model are the same as for that with the SFDM halo, the same mass for the disk and the same mass enclosed at a radius $r_0$, this parameters fixed by observations.
The best fit with the NFW + disk mass model has parameters $a=3$ \rm{$kpc$}, $b=0.3$ \rm{$kpc$} for the disk and $r_s=10$ \rm{$kpc$} for the dark matter halo. With a reduced chi-square $\chi_r^2=1.5$ the fit is good but not better than that with the cored halo.

In the figures we include also the analytic rotation curves computed from each model in order to illustrate the contribution of the live gas to the fit of the data.

\begin{figure}
\begin{center}
\includegraphics[width=8.7cm]{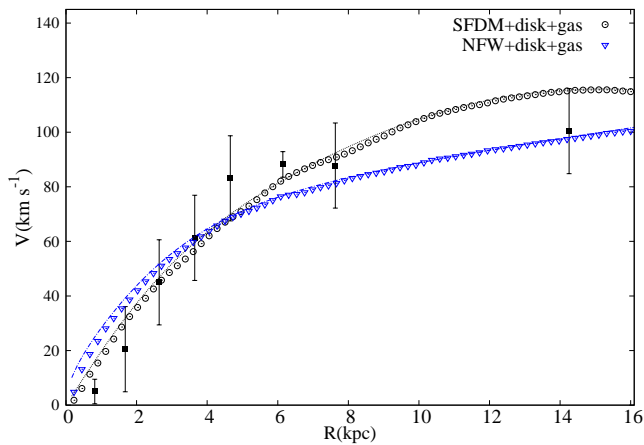}
\end{center}
\caption{Lsb galaxy F568-3: Compared Rotation curve from data (square dots), simulation with the SFDM halo + disk (open circles), simulation with the NFW halo + disk (triangles), and the initial rotation curve for both models (dashed lines).}
\label{fig:F5683}
\end{figure}

F5683: For this galaxy we used the parameters $a=1.3$ \rm{$kpc$}, $b=0.2$ \rm{$kpc$} for the disk and $a_h=5.2$ \rm{$kpc$} for the SFDM halo. Figure \ref{fig:F5683} still shows a relatively good fit to the data, $\chi_r^2=1.8$
Here the NFW + disk mass model that fits better the data with parameters $a=5$ \rm{$kpc$}, $b=0.3$ \rm{$kpc$} for the disk and $r_s=200$ \rm{$kpc$} for the dark matter halo, and a reduced chi-square $\chi_r^2=5.03$ has not a very good fit.

The mass model with a cored halo is more favored by the data, as we can see in Figure \ref{fig:F5683} that the cuspy halo fails in recovering the inner points of the rotation curve.

\begin{figure}
\begin{center}
\includegraphics[width=8.7cm]{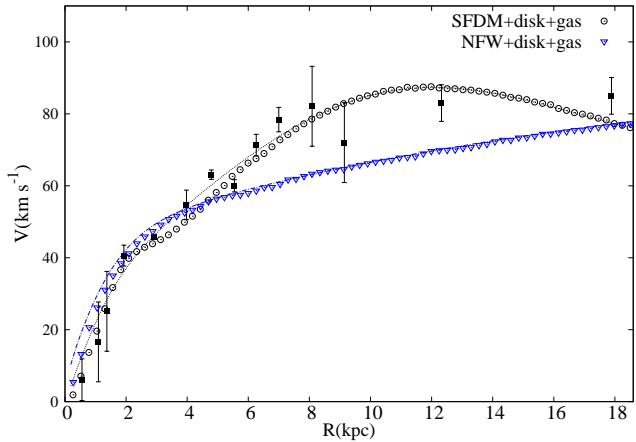}
\end{center}
\caption{Lsb galaxy F583-1: Compared Rotation curve from data (square dots), simulation with the SFDM halo + disk (open circles), simulation with the NFW halo + disk (triangles), and the initial rotation curve for both models (dashed lines).}
\label{fig:F5831}
\end{figure}

F5831: The cored mass model for this Lsb galaxy fits better the data with parameters $a=1.3$ \rm{$kpc$}, $b=0.2$ \rm{$kpc$} for the disk and $a_h=5.2$ \rm{$kpc$} for the dark matter halo. Figure \ref{fig:F5831} shows a relatively good fit, $\chi_r^2=4.8$, and the simulation rotation curve recovers well the innermost part of the data.
The best fit to the data for the cuspy mass model has parameters $a=2.5$ \rm{$kpc$}, $b=0.15$ \rm{$kpc$} for the disk and $r_s=45$ \rm{$kpc$} for the dark matter halo and reduced chi-square $\chi_r^2=9.5$. Although Figure \ref{fig:F5831} shows the best fit for the cuspy mass model, its rotation curve is not able to meet the inner and outer points of the data at the same time. The fit in the inner region of the rotation curve is almost as good as the cored model but misses the rest of the data.

\begin{figure}
\begin{center}
\includegraphics[width=8.7cm]{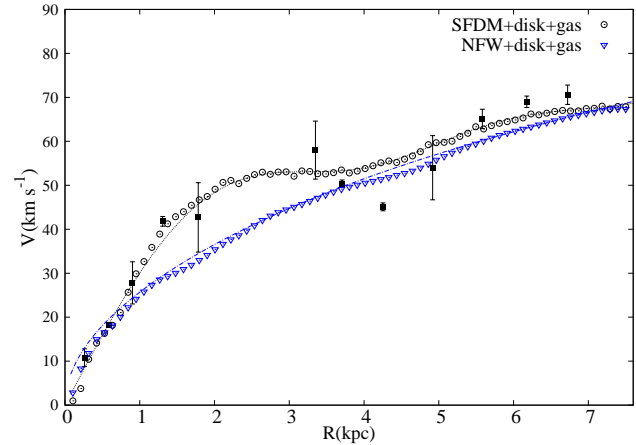}
\end{center}
\caption{Lsb galaxy F583-4: Compared Rotation curve from data (square dots), simulation with the SFDM halo + disk (open circles), simulation with the NFW halo + disk (triangles), and the initial rotation curve for both models (dashed lines).}
\label{fig:F5834}
\end{figure}

Finally, the data for the lsb galaxy F583-4 is lest smooth, with a bump in the rotation curve. Is hard to fit with a smooth halo, but our mass model still recovers the most of the data points with  $\chi_r^2=21.4$.
In contrast we see in Figure \ref{fig:F5834} that because of the universality of the rotation curve profile for a NFW halo, is more difficult to fit data like these.
The best fit to the data for the cuspy mass model has a a reduced chi-square $\chi_r^2=26.1$ that do not improves that of the cored mass model.

Table \ref{table:table1} summarise the parameters used in our simulations with the cored mass model. The total mass ($M_{total}$), disk mass ($M_{disk}$) and HI mass ($M_{HI}$) are fixed by observations \citep{K11b}. The scale-lengths of the disk ($a$, $b$) and the halo ($a_h$) span a range of values in order to obtain good fits.

\begin{table*}\centering
\ra{1.3}
\begin{tabular}{@{}lccccccc@{}}\toprule
       & $M_{total}$        & $M_{disk}$         & $M_{HI}$           & \phantom{ab}& $a$   & $b$   & $a_h$\\
       &$(10^{10}\rm{M_{\odot}})$&$(10^{10}\rm{M_{\odot}})$&$(10^{10}\rm{M_{\odot}})$& \phantom{ab}&\rm{$(kpc)$}&\rm{$(kpc)$}&\rm{$(kpc)$}\\
\midrule
F579-V1&   4.3              &     0.63           &    0.11            &              & 1.3   & 0.2    & 5.2  \\
F568-3 &   5.6              &     0.8            &    0.39            &              & 4.5   & 0.5    & 7  \\
F583-1 &   2.5              &     0.185          &    0.16            &              & 2.5   & 0.15   & 5.5\\
F583-4 &   0.76             &     0.077          &    0.077           &              & 1.2   & 0.1    & 1.2, 7\\
\bottomrule
\end{tabular}
\caption{Parameters used in the simulations and mass models. Total dynamical mass $M_{total}$, baryonic disk mass $M_{disk}$, HI mass $M_{HI}$, disk scale-lengths $a$, $b$ and DM halo scale-length $a_h$.}
\label{table:table1}
\end{table*}

\citet{R13} use a combination of states to fit rotations curves that spans at large radii and are not smooth. In this work, for the SFDM halo, we used only one state in three of the four galaxies, while F583-4 needed two states. So for this three galaxies the number of free parameters was the same for the models compared here. The added parameter in F583-4 is taken into account when computing the reduced chi-square by penalizing for this extra parameter.

It is important to note that the rotation velocity for the gas deviates just a little from the analytical circular velocity for the two models compared here. The bumps and wiggles seen in the simulations with the SFDM halo are driven almost entirely for the mass model with a small effect due to the gas.

\section{Summary and Conclusions}\label{conclusions}

The Scalar field has been proposed as a galactic dark matter model and has been tested by confronting with rotation curve data. It has been shown that when using the minimum disk hypothesis (neglecting the baryonic component) in halos with one or more nodes in their density profiles, fits to rotation curves for LSB galaxies improve significantly.
Adding the baryonic component to the mass model results in a better fit, but including live gas is a step forward from the previous works using SFDM, as for example, the rotation velocity of the gas is not always exactly equal to the circular velocity of a test particle.

LSB galaxies are dark matter dominated and when building a mass model for this objects, a cored DM halo is favored by the data. When comparing the cored and the cuspy mass models, we see that a NFW halo does it relatively well in some cases, but a SFDM halo does it better for this kind of galaxies, specially in the inner regions.

There are differences with previous works with scalar field at zero temperature, considering a finite temperature for the scalar field configuration is translated in a non-universality for the rotation curves profile of the mass model, this allows us to describe well different lsb galaxies. As already showed in previous works with this density profile, the non-universality of the rotation curves allows also to account for wiggles and bumps in the velocities data. We found that this features are too wide to be addressed to the physics of the baryons, as modeled here, and can be explained within the SFDM model, being inherent to the DM halo.

The initial conditions presented here have been tested against dynamical stability ensuring that the obtained rotation curves profiles do not present artificial effects.

We have performed three-dimensional hydrodynamic simulations with simple physical processes for the gas (we do not model star formation or feedback) that evolves within a static dark matter and stellar disk potential. And although we argue that huge quantities of feedback events (star formation, supernovas) are not needed, those should be added to our simulations but in a way different to that used in CDM-based simulations because we are not trying to change the shape of the DM density profiles through feedback processes.

The results presented here show that the observed bumps and wiggles in the rotation curves can be explained by the dark matter halo, and that gas motions as modeled here are not sufficient to explain this features. But further work is required by considering a live dark matter halo that responds to the gaseous and stellar component, this becomes important for long term evolution of a bar or a spiral pattern that might also explain the observed bumps and wiggles.

\section*{Acknowledgments}
The authors wish to thank to Victor H. Robles for many helpful discussions and to the anonymous referee for
helping us to improve the manuscript.
The numerical computations were carried out in the "Laboratorio de Super-C\'omputo Astrof\'{\i}sico (LaSumA) del Cinvestav".
The authors acknowledge to the General Coordination of Information and 
Communications Technologies (CGSTIC) at CINVESTAV for providing HPC 
resources on the Hybrid Cluster Supercomputer "Xiuhcoatl".
This work was partially supported by CONACyT M\'exico under grants CB-2009-01, no. 132400, CB-2011, no. 166212,  and I0101/131/07 C-234/07 of the Instituto Avanzado de Cosmologia (IAC) collaboration
(http://www.iac.edu.mx/). LAMM is supported by a CONACYT scholarship.

\label{lastpage}

\end{document}